\documentclass[12pt,epsf]{article}
\usepackage{graphicx, amsmath}

\begin{document}

\sloppy
\begin{flushright}{SIT-HEP/TM-27}
\end{flushright}
\vskip 1.5 truecm
\centerline{\large{\bf Primordial black holes from
cosmic necklaces}} 
\vskip .75 truecm
\centerline{\bf Tomohiro Matsuda
\footnote{matsuda@sit.ac.jp}}
\vskip .4 truecm
\centerline {\it Laboratory of Physics, Saitama Institute of
 Technology,}
\centerline {\it Fusaiji, Okabe-machi, Saitama 369-0293, 
Japan}
\vskip 1. truecm
\makeatletter
\@addtoreset{equation}{section}
\def\theequation{\thesection.\arabic{equation}}
\makeatother
\vskip 1. truecm

\begin{abstract}
\hspace*{\parindent}
Cosmic necklaces are hybrid topological defects consisting of monopoles
 and strings.
We argue that primordial black
 holes(PBHs) may have formed from loops of the necklaces, if there exist
 stable winding states, such as coils and cycloops. 
Unlike the standard scenario of PBH formation from string loops,
in which the kinetic energy plays important role when strings
collapse into black holes, the PBH formation may occur in our scenario
after
necklaces have dissipated their kinetic energy.
Then, the significant difference appears in the
 production ratio.
In the standard scenario, the production ratio $f$ becomes a
 tiny fraction $f\sim 10^{-20}$, however it becomes $f \sim 1$
in our case.
On the other hand, the typical mass of the PBHs is much smaller than
 the standard scenario, if they are produced in the same epoch.
As the two mechanisms may work at the same time, the necklaces may have
 more than one channel of the gravitational collapse.
Although the result obtained in this paper depends on the evolution
of the dimensionless parameter $r$, the existence of the
 winding state could be a serious problem in some cases.
Since the existence of the winding state in brane models is due to the
 existence of a 
 non-tivial circle in the compactified space, the PBH formation can be
used to probe the structure of the compactified space.
Black holes produced by this mechanism may have peculiar properties.

\end{abstract}

\newpage
\section{Introduction}
Cosmic strings have recently gained a great interest. 
In the context of a brane world scenario, cosmic strings can be produced
after brane inflation\cite{brane-inflation0, angled-inflation}.
It has been discussed that such strings
lead to observational predictions that might be used to distinguish 
brane world from conventional models\cite{vilenkin-lowp}.
From phenomenological viewpoints, the idea of large extra
dimension\cite{Extra_1} is important for such 
models, because it may solve the hierarchy problem.
In the scenarios of large extra dimension, the fields in the standard
model are localized on wall-like structures, while
the graviton propagates in the bulk. 
In the context of string theory, a natural embedding of
this picture is realized by brane construction.
The brane models are therefore interesting both from phenomenological
and cosmological viewpoints.
In order to find signatures of branes, analysis of cosmological defect
formation and evolution is important.\footnote{Inflation in models of
low fundamental scale   
are discussed in ref.\cite{low_inflation, matsuda_nontach,
matsuda_defectinfla}.
Scenarios of baryogenesis in such models are discussed in ref.
\cite{low_baryo, Defect-baryo-largeextra, 
Defect-baryo-4D}, where defects play distinguishable roles.}
Brane defects such as monopoles, strings, domain walls and
Q-balls are discussed in ref.\cite{BraneQball,
matsuda_monopoles_and_walls, incidental, matsuda_angleddefect,
matsuda_necklace,overproduction}, where it has been 
concluded that not only strings but also other defects should
appear.

The purpose of this paper is to find distinguishable property of
non-standard strings focusing 
our attention on the production of primordial black holes.
S.W.Hawking has discussed\cite{Hawking} that cosmic string loops
that shrink by a factor of order $1/G\mu$ will form black
holes.\footnote{See ref.\cite{other_PBH} for other possible origin and
effects of PBHs.} 
He estimated that a fraction $f$ of order $(G\mu)^{2x-4}$ loops will form
black holes, where $x$ is a ratio of the loop length to the correlation
length.
The result obtained in ref.\cite{Hawking} is cosmologically important
because the emission of $\gamma$-rays from such little black holes could
be significant\cite{post-Hawking}. 
Numerical simulation of the loop fragmentation and evolution is obtained
in ref.\cite{PBH_numerical}, where the fraction $f$ becomes
\begin{equation}
f\simeq 10^{5}\times(G\mu)^{4}.
\end{equation}
Black holes created by this collision are so small that they lose their
energy due to the Hawking evaporation process.
The fraction of the critical density of the Universe in primordial black
holes(PBHs) today due to collapsing cosmic string loops is discussed in
ref.\cite{branden};
\begin{equation}
\Omega_{PBH}(t_0) = \frac{1}{\rho_{crit}(t_0)}\int^{t_0}_{t_*}
dt \frac{dn_{BH}}{dt}m(t,t_0).
\end{equation}
Here $t_0$ is the present age of the Universe, and $t_*$ is the time
when PBHs with initial mass $M_*=4.4\times 10^{14}g$ are formed,
which are expiring today.
$m(t,t_0)$ is the present mass of a PBH created at time $t$, which can
be approximated as $m(t,t_0))\simeq \alpha \mu t$.
The extragalactic $\gamma$-ray flux observed
at 100MeV is commonly known to provide a strong constraint on the
population of black holes today.
According to ref.\cite{Gibbons-carr}, the limit implied by the EGRET
experiment becomes
\begin{equation}
\label{PBH_omega}
\Omega_{PBH} < 10^{-9}.
\end{equation}
The scaling solution of the conventional string network suggests that the
rate of PBH formation is
\begin{equation}
\label{string_BH}
\frac{d n_{BH}}{dt} = f\frac{n_{loop}}{dt}\sim \alpha^{-1}f t^{-4}.
\end{equation}
Using the above results one can obtain an upper bound\cite{branden}
\begin{equation}
G\mu <  10^{-6},
\end{equation}
which is close to the constraint obtained from the normalization of the
cosmic string model to the CMB.
Based on the above arguments, we have analyzed in our previous paper 
the PBH formation in less simplified system\cite{matsuda_pbh1}. 
In ref.\cite{matsuda_pbh1} we have considered two examples;
Monopole-antimonopole pairs connected by strings and monopole-string
network of the $Z_n ~(n>2)$ strings.
The latter is a model of monopole-string network, in which monopoles
are connected to $n>2$ strings.

Considering the above arguments, it is
conceivable that primordial black holes can generally form from both 
standard string loops and monopole-string networks.\footnote{
Here we should note that the PBH production from necklaces
is different from the $Z_n$ strings that we have discussed in
\cite{matsuda_pbh1} both in qualitative and quantitative properties. 
The mechanism that we will consider in  
this paper is quite different from the scenarios that we have discussed
in ref.\cite{matsuda_pbh1}.
In the Hawking's scenario\cite{Hawking} and in ref.\cite{matsuda_pbh1},
the kinetic energy of the collapsing objects(strings or monopoles) play
important role.
In this case, the gravitational collapse is due to their huge kinetic
energy when they collide.
On the other hand, the kinetic energy play no role in this paper.
This is the crucial difference between our present scenario and the
previous scenarios of PBH formation from cosmic defects.} 
In this paper, we consider a more complexified model of monopole-string
network; cosmic necklaces.
Berezinsky and Vilenkin\cite{vilenkin-necklace} found that if one
started with a low density of monopoles $r\ll 1$, where $r$ is the ratio
of monopole energy density to string energy density per unit length, one
can approximate 
the evolution of the system by the standard evolution of a string network,
and also if one could ignore monopole-antimonopole annihilation, the
density of monopoles on strings would increase until the point where
the conventional-string approximation breaks down.
In ref.\cite{vilenkin-necklace} the authors leave the detailed analysis
of the evolution of such systems to numerical simulations, in particular
the effect of monopole-antimonopole annihilation.
Later in ref.\cite{necklace-simulation} numerical simulations of cosmic
necklaces are performed, in which it has been found that the string motion
is periodic when the total monopole energy is much smaller than the 
string energy, and also that the monopoles travel along the string and
annihilate with each other.
In this paper, we consider networks of necklaces where
monopole-antimonopole annihilation is not suppressed.
We can therefore approximate the evolution of necklaces by the standard
evolution of simple strings, at least during the era when significant
amount of primordial black holes are produced.

\begin{figure}[ht]
 \begin{center}
\begin{picture}(400,420)(0,0)
\resizebox{10cm}{!}{\includegraphics{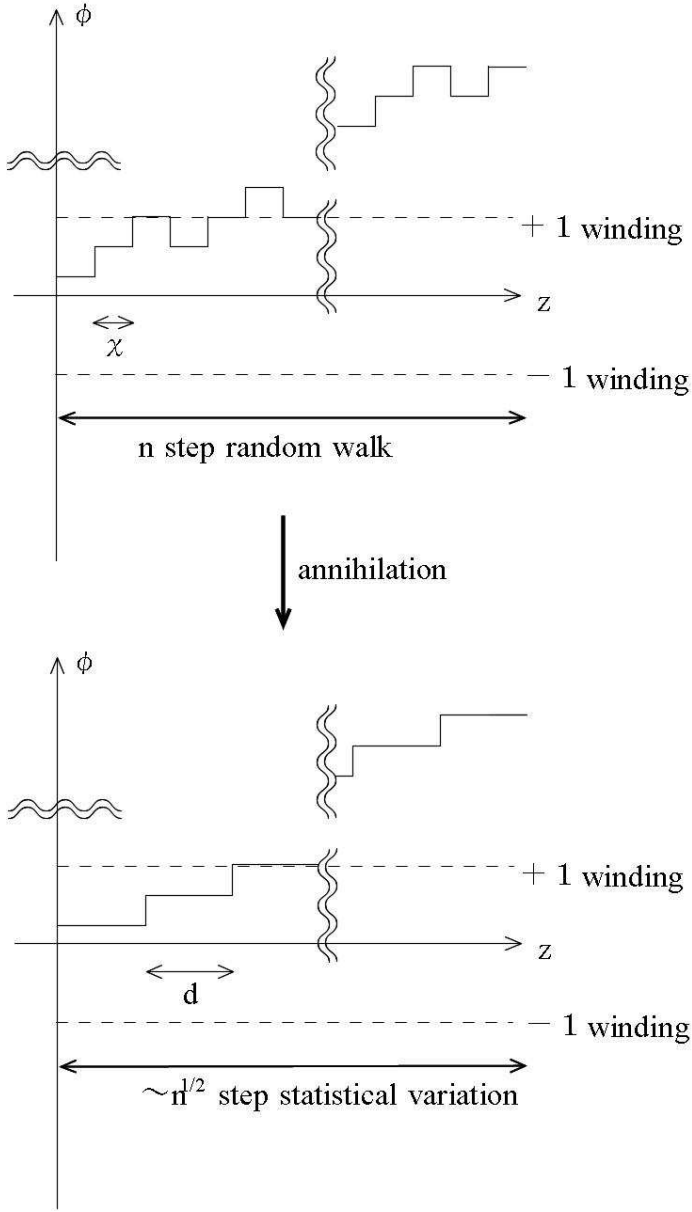}} 
\end{picture}
\caption{The figure in the first line shows the $n$-step random walk
  on a string. 
  Each kink(antikink) corresponds to the left(right) mover.
  Here the step length between each random walk is denoted by $\chi$.
  In the case that the annihilation of the kink-antikinks is
  efficient, about $\sim n^{1/2}$ kinks(antikinks) will remain due to
  the statistical variation.
  The distance between kinks that remain after efficient annihilation is
  denoted by $d$.
  In our paper, we assume that the evolution of $\chi$ 
  is given by $\chi \propto t^{k-1}$, where the precise value of $k$
  could be determined by numerical simulations.}
\label{fig:random}
 \end{center}
\end{figure}

In the case that the winding states of necklaces are stabilized, the
simple statistical argument of a random walk indicates that about
$n^{1/2}$ 
of the initial $n$ monopoles on a long string could survive,
even if monopole-antimonopole annihilation is an efficient
process.\footnote{See fig.\ref{fig:random}}
If loops have formed from such strings, heavy winding states (coils)
remain\cite{matsuda_necklace}.
In this case, one can expect that the nucleating rate of such winding
states decreases, while the mass increases with time.
A similar argument has been discussed in ref.\cite{cycloops} where
the authors have assumed that the step length between random
walk($\chi(t)$)  is a constant that does not depend on time.
On the basis of this assumption, they have obtained the time-dependence
of the 
mass of the winding states(cycloops), $m_{cycloops} \propto t^{1/2}$.
One might think that it should be appropriate to assume that $\chi$
``increase'' with time due to the expansion of the Universe.
However, considering the result obtained in
ref.\cite{vilenkin-necklace} it is obvious that one cannot simply 
ignore the possibility that $\chi$
``decreases'' with time.
Obviously, this possibility cannot be ignored even in the case that 
 the actual distance
between monopoles increases due to annihilation.
In this case, efficient annihilation only reduces the number of monopoles
from $n$ to $\sqrt{n}$.\footnote{$\chi$ is much smaller than the actual
distance between monopoles, $\chi \ll d$.}
In this paper, we therefore assume that $\chi$
depends on time as
\begin{equation}
\label{dk}
\chi(t) \propto t^{k-1},
\end{equation}
where $k\simeq 0$ corresponds to $\kappa_g-\kappa_s\simeq 1$ in
ref.\cite{vilenkin-necklace}.
Since the mass of the winding state always increases with time, 
one cannot ignore the possibility that the coils could turn into
black holes.
Of course, Hawking's mechanism of black hole
formation from string loops\cite{Hawking} works simultaneously.
We will show that in the networks of necklaces-coils the old mechanism
of PBH formation is less efficient than the new mechanism.

\section{Necklaces stabilized by windings}
We consider necklaces whose loops are stabilized by
windings around a non-trivial circle in their moduli space.
A coil could be a higher-dimensional object(brane) that winds
around a non-trivial circle in the compactified space, or could be
a non-abelian string that has a non-trivial circle in their 
moduli space.
As we have stated above, the simple
statistical argument of a random walk indicates that about $n^{1/2}$
of initial $n$ monopoles on a long string could survive
when winding states of necklaces are stabilized,
even if monopole-antimonopole annihilation is an efficient process. 
If loops have formed from such strings, heavy winding states (coils)
would remain\cite{matsuda_necklace}.
Here we show two concrete examples.
The first is an example of cosmic string produced after brane
inflation, and the second is an example of non-abelian necklace.
 
\subsection{Necklaces produced after angled inflation}
It is sometimes discussed that only cosmic strings are produced after
brane inflation.
However, the argument is often based on the assumptions;
\begin{itemize}
\item There is only a pair of $D\overline{D}$ brane in the inflation
      sector. 
\item There is no ``overproduction''\cite{overproduction}
\item There are no local vacua in the moduli space of such strings
      so that no kinks appear on the strings.
      One should also assume that strings cannot move and 
      wind around a non-trivial circle in the compactified space.
\end{itemize}
However, as we have discussed in ref.\cite{BraneQball,
matsuda_monopoles_and_walls,matsuda_necklace}, defects other than
strings can be produced in generic models of brane inflation. 
Strings turn into necklaces\cite{matsuda_necklace} when there exists
(quasi-)degenerated local vacua in their moduli space.  
In this case, kinks appear on strings interpolating
between domains of (quasi-)degenerated vacua.
One can also consider windings around internal space, which is called
``coils'' in ref.\cite{matsuda_necklace} and later named ``cycloops'' in 
ref.\cite{cycloops}.
Perhaps the simplest scenario of $D\overline{D}$ annihilation is
rather problematic if one take seriously the mechanism of
reheating\cite{brane-reheat}.
It is
therefore important\cite{matsuda_angleddefect, matsuda_necklace}
 to consider less simplified model of brane inflation.
For example, angled inflation is a model of brane inflation in which
inflating branes does not merely 
 annihilate but reconnect at the end of inflation.
In this case the daughter branes are not free propagating in the bulk,
but have their endpoints on their mother branes. 
To make our discussions simple and convincing, here we consider angled
brane inflation with a small angle($\theta \ll
1$)\cite{matsuda_angleddefect, matsuda_necklace}. 
In angled inflation, cosmic strings are the daughter $D_{p-2}$ branes
extended between mother $D_p$ branes. 
The $D_{p-2}$ branes have the flat direction (moduli) in the compactified
space, along which the endpoints of the $D_{p-2}$ branes can move
 freely on the $D_p$ branes.
Then the position of the $D_{p-2}$ branes can vary along a
cosmic string, which results in $(1+1)$-dimensional kink configurations
appearing on the string.
One can also find a winding state that is defined on the worldvolume
 of the strings\cite{matsuda_necklace}. 
In this case, kinks that appear on strings are monopoles, and 
they are produced by spatial deformation of the $D_{p-2}$ branes.
Therefore, the brane necklaces are the {\bf hybrid} of
 ``brane creation'' and  ``brane deformation''.
The chopped loops of the necklaces can shrink to produce stable
winding states, which look like coils winding around compactified space.
It will be helpful to note here that a brane coil winds around the
compactified space that is {\bf different} from the mother branes.
Of course the mechanism that induces such windings is different from 
the conventional Kibble mechanism, therefore our argument does not
contradict to the 
previous arguments\cite{previous-onlystrings} that have suggested only
strings are produced after 
brane inflation.

\subsection{Non-abelian necklaces}
Besides the brane-induced necklaces that we have discussed above, 
one can construct another model that have a similar configuration
without invoking brane dynamics in compactified space.
In this case, the flat direction in the compactified space is
replaced by a moduli space of the worldvolume effective action of 
a non-abelian string.
Let us consider the dynamics of cosmic strings living in a non-abelian
$U(N_c)$ gauge theory that is coupled to $N_f$ scalar fields $q_i$,
which transform in the fundamental representation;
\begin{equation}
L = \frac{1}{4e^2}Tr F_{\mu\nu}F^{\mu\nu} + \sum^{N_f}_{i=1}
{\cal D}_{\mu}q_i^{\dagger}{\cal D}_{\mu}q_i^{\dagger}
-\frac{\lambda e^2}{2}\left(\sum^{N_f}_{i=1} q_i \otimes q_i^{\dagger}
-v^2\right)^2.
\end{equation}
The above Lagrangian has a $SU(N_f)$ flavor symmetry, which rotates the
scalars. 
We can also include explicit symmetry breaking terms into the
Lagrangian, which breaks global flavor symmetry.
The most obvious example is a small mass term for the scalars;
\begin{equation}
V_{br1}\sim \sum_{i}m_i^2 q_i^{\dagger}q_i,
\end{equation}
which shifts the vacuum expectation value to\cite{tong_hashimoto}
\begin{equation}
q^a_i = \left(v^2-\frac{m^2_i}{\lambda e^2}\right)^{1/2}\delta^a_i.
\end{equation}
Then one may embed an abelian vortex in the i-th $U(1)$ subgroup
of $U(N_c)$, whose tension becomes $T_i \sim
\left(v^2-\frac{m^2_i}{\lambda e^2}\right)^{1/2}$.
In this case, due to the difference between the string tension,
monopoles could form binding states.

One may extend the model to $N=2$ supersymmetric QCD 
or simply include an additional adjoint scalar field $\phi$.
In any case, the typical potential for the adjoint scalar could
be\cite{tong_hashimoto} 
\begin{equation}
V_{br2}\sim \sum_{i}q_i^\dagger(\phi-m_i)^2 q_i.
\end{equation}
Here the potential breaks $U(1)_R$ symmetry.
The tensions of the strings degenerate in this case.

Alternatively, one can consider supersymmetry-breaking potential
that could be induced by higher-dimensional effects,
\begin{equation}
V_{br3}\sim \sum_{i}q_i^\dagger(|\phi|^2-m^2) q_i,
\end{equation}
which preserves $U(1)_R$ symmetry.
In this case, due to D-flatness condition if supersymmetry is imposed,
the vacuum expectation value of the adjoint field is placed on a circle
and given by\cite{matsuda_monopoles_and_walls}
\begin{equation}
\phi=m\times  
diag(1,e^{\frac{2\pi}{N_c}}, e^{\frac{2\pi}{N_c}\times 2},
...e^{\frac{2\pi}{N_c}\times (N_c-1)}).
\end{equation}
One can break the remaining classical $U(1)_R$ symmetry by adding an
explicit breaking term, or by anomaly\cite{matsuda_monopoles_and_walls,
tong_hashimoto}.  

In any case,  strings living in different $U(1)$ subgroups can transmute each
other by the kinks(walls on their 2D worldvolume) that interpolate
between degenerated vacua. 

Is it possible to construct ``winding'' state
from non-abelian necklaces, which look like coils in 
brane models? 
A similar argument has already been discussed by Dvali, Tavartkiladze and
Nanobashvili\cite{winding_wall} for $Z_2$ domain wall in
four-dimensional theory.
The authors have discussed that ``windings'' may stabilize the
wall-antiwall bound state if the potential is steep in the radial
direction.
Of course, a similar situation may happen if (for example) the origin is
lifted by an effective potential $\sim \phi^{-n}$. 
The important point here is whether the absolute value of the
scalar field vanishes inside the bound state of walls(kinks).
If  ``windings'' of such kinks cannot be resolved due to the potential
barrier near the origin, naive annihilation process is inhibited and
stable bound state will remain.
The same ``windings'' may happen in our case, where kinks are 
corresponding to domain walls on (1+1)-dimensional world.
The winding states are therefore generic remnants of
non-abelian necklaces.

\subsection{Distance between monopoles}
In this paper, we assume that monopole-antimonopole annihilation
is efficient so that $r \ll 1$ is a good approximation.
Since we are considering a model in which the winding states of
necklaces are stabilized, the simple 
statistical argument of a random walk indicates that about $n^{1/2}$
of the initial $n$ monopoles on a long string could survive
even if monopole-antimonopole annihilation is an efficient process.
If loops have formed from such strings, stable winding states would
remain.
In any case, the initial distance between monopoles is important.
Typically the number density to entropy ratio is expressed by the
standard formula 
\begin{equation}
\frac{n_m}{s} >
\left[\left(\frac{T_M}{M_p}\right)ln\left(\frac{M_p^4}{T_M^4}
\right)\right]^3.
\end{equation}
Therefore, distance between monopoles when necklaces are formed at
the temperature $T=T_n$(or at the time $t=t_n$) is bounded
by\cite{vilenkin_book} 
\begin{equation}
 d(t_n) < (t_n t_M)^{1/2} \sim T_n^{-1} \left(\frac{M_p}{T_M} \right),
\end{equation}
which gives us the {\bf maximum} value of $\chi$
at $t=t_n$.\footnote{In brane models, if monopole production occurs
later than the string production, what appears during the period between
string production and $t_n$ is called cycloops\cite{cycloops}.
After $t_n$, when the lift of the potential becomes significant,
cycloops turns into necklaces on which kinks(beads) are interpolating
between vacua.}

\subsection{Mass of coils}
Let us assume that a loop of the length $l(t)$ initially contains
$n$ monopoles.
Then from the simple statistical argument of a random walk 
one can obtain the mass of the stable relic;
\begin{equation}
M_{coil}(t) \sim n(t)^{1/2} m.
\end{equation}
Here we assume that the number of monopoles that are initially contained
in a loop is given by
\begin{equation}
\label{number-monopole_ini}
n(t) \sim \frac{l(t)}{d(t_n)\times \left(\frac{t}{t_n}\right)^{k-1}}.
\end{equation}
For example, in the case when $k=0$,
the initial number of monopoles that could be contained in a loop that
is just chopped off from the string networks is  
\begin{equation}
\label{number-monopole}
n(t) \sim \frac{l(t)}{d(t_n)\times \left(\frac{t}{t_n}\right)^{-1}}.
\end{equation}

It will be helpful to comment about the crucial points 
related to the time-dependence of the mass of the 
winding states.
\begin{enumerate}
\item
In the standard scenario\cite{Hawking}, it has been discussed that
the mass of PBHs increases with time as $m_{PBH}\sim \alpha \mu t$,
while the production rate $f\sim 10^{-20}$ is a tiny fraction.
In our case, the mass of PBHs depends on time as
\begin{equation}
m_{PBH} \propto t^{\frac{2-k}{2}}
\end{equation}
where $k$ may have gap at $t=t_{eq}$.
In our case, the mass of PBHs could be either small($k<0$) or large($k>0$) 
compared to the original scenario.
The bound in our model is therefore quite sensitive to the value of $k$,
which suggests that numerical simulations are quite important for
     further study.
\item
The present number density of black holes with mass $M$
can be obtained by redshifting the distribution from the time of their
formation to the present time $t_0$.
Then we can calculate $d n_{BH}/dM$.
For example, let us consider $k=\delta$ when there is a small 
($\delta \ll 1$) deviation from $k=0$.
Then, we have
\begin{equation}
\frac{d n_{BH}(M)}{d M} \propto M^{-2.5-0.75\delta}.
\end{equation}
Obviously, deviation from the standard mass-dependence is a unique
     property of our scenario.
This can be used to distinguish conventional strings from
     necklaces/coils.  
\end{enumerate}

\subsection{Loops from scaling necklaces}
Let us assume that PBH formation starts after friction-domination has
been ended at $t=t_{scale}$.
In this case, one can assume that the typical velocity of necklaces 
is $v\sim 1$.
If the monopoles does not have any unconfined magnetic charge,
one can assume that necklaces scale and the number of loops produced
from the network is
\begin{equation}
\label{loop-production}
\frac{d n(t)_l}{dt} \sim \frac{1}{\alpha N^{-1} t^4},
\end{equation}
where $N$ is the number of (quasi)-degenerated vacua of the necklace.
Our result is based on the assumption that the evolution of the network
of the necklaces
is identical to the strings of low reconnection ratio; $p \sim
N^{-1} \ll 1$\cite{matsuda_necklace}.

\subsection{PBHs from necklaces}
\begin{figure}[ht]
 \begin{center}
\begin{picture}(400,280)(0,0)
\resizebox{9cm}{!}{\includegraphics{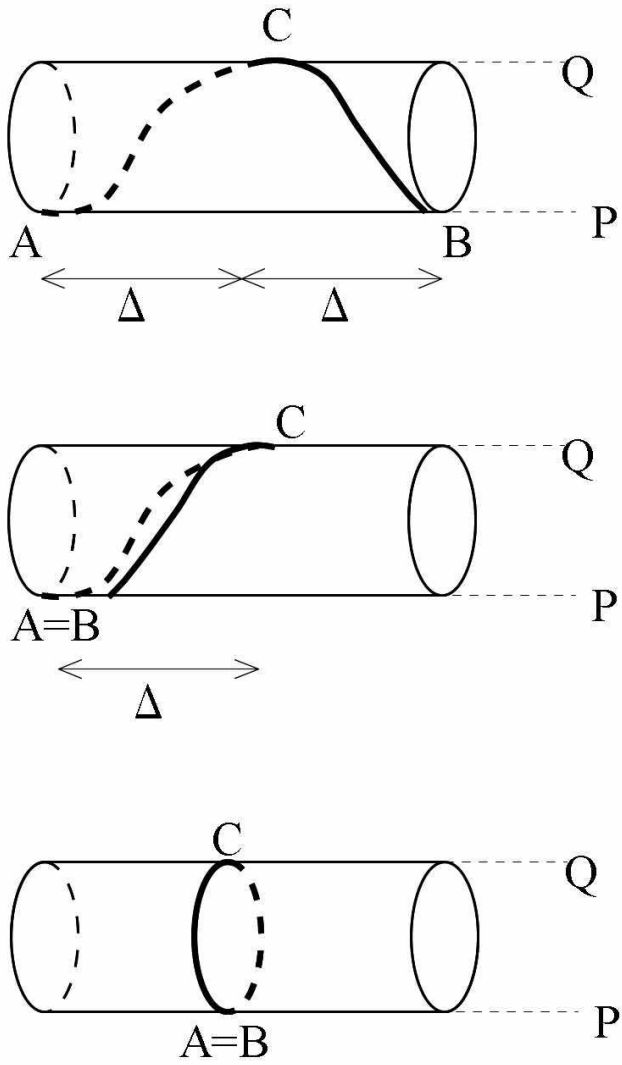}} 
\end{picture}
\caption{The figure in the first line shows the winding that appears on
  a long necklace. Here the width of the kinks $A-C$ and $C-B$ is denoted
  by $\Delta$. There are degenerated vacua at $A, B$
  and $C$ that appear on the necklaces. 
  The figure in the middle shows a loop that is formed
  by the necklace. In this case, the left end is identified with the
  right.
  The figure in the last line shows the conventional winding state
  of the brane. In general, the tension of the strings are much
  larger than the square of the effective mass that lifts the flat
  directions.
  Then, the width of the winding
  state becomes about the same order as the width of the brane itself,
  as is shown in the last line.
  As we are not considering any peculiar repulsive force acting between
  the $D$-branes, the width of the n-winding state cannot be
  proportional to n. }
\label{fig1}
 \end{center}
\end{figure}

As we have discussed above, mass of the coils tends to increase with
time.
In our case, although the typical length of loops increases as $l\propto
t$, the mass of the coils depends on time as $m_{coil}\propto
t^{\frac{2-k}{2}}$.
Initially the motion of necklaces is damped due to frictional
force until $t_{scale} \sim (G\mu)^{-1}t_s$, where $t_s$ is the time of
string 
formation.
After damped epoch($t>t_{scale}$), network of necklaces starts
scaling.\footnote{Although the constant $k$ could have a gap at
$t=t_{scale}$, here we assume that the gap is negligible.}
Although it might depend on the form of the potential that induces kinks
on the strings, $k_n$ kinks(monopoles) on long string will
have width $\delta_k > k_n\times \delta_m$, where $\delta_m$
denotes the width of a monopole.
One can therefore assume that monopoles cannot turn into PBHs as far as
they are moving on a long string.
A chopped string form loops that can shrink to a point (at least in the
uncompactified space) after they have
dissipated their kinetic energy.\footnote{If one wants to discuss PBHs
from loops {\bf before} dissipation, the ratio becomes
tiny\cite{Hawking}. Note that what we are considering here is different
from the original scenario\cite{Hawking}.}
Even in the case that the width of a kink is wide due to a shallow
potential in the 
moduli space, these kinks can turn
into a point-like winding state.
Here we do not discuss about the details of the condition, but simply
assume that we are considering a model in which loops can shrink to a
point in the four-dimensional space due to their 
string tension.\footnote{See fig.\ref{fig1}.}

Now we can estimate the time when coils start to turn into black holes.
The Schwarzschild radius of a loop which contained initially $n$
monopoles is 
\begin{equation}
\label{sch-coil}
R_g\sim \sqrt{n(t)} m/M_p^2,
\end{equation}
where $m$ is the mass of a monopole.
In the case when a loop shrinks to a point-like state whose
width is $\delta_{coil}\sim \eta_s^{-1}$,
the condition for PBH formation becomes
\begin{equation}
\label{pbh-coil}
R_g > \eta_s^{-1}.
\end{equation}
Using eq.(\ref{number-monopole_ini}),(\ref{sch-coil}) and
(\ref{pbh-coil}), 
we can calculate $t_{PBH}$ when PBH formation starts;
\begin{equation}
t_{PBH}\sim 
\left(\frac{M_p^4 d(t_n)t_n^{1-k}}{\eta_s^2 m^2 \alpha}\right)^{1/(2-k)}.
\end{equation}
To calculate the present density of primordial black holes,
we should consider black holes whose mass is larger than $M_*$, as we
have discussed in section 1.
In the natural evolution ($k=0$), one can easily obtain a severe bound for
the string 
tension\cite{branden}.
In this case, the explicit form of the mass is
\begin{equation}
m_{coil} \sim \alpha^{1/2} (d(t_n) t_n)^{-1/2} m t.
\end{equation}
From eq.(\ref{loop-production}), one can obtain 
\begin{equation}
G\mu < 10^{-20} \times 
\left[\frac{p}{10^{-2}}\right]^{4/5}
\left[\frac{\gamma}{10^{2}}\right]^{1/5}
\left[\frac{t_n}{M_p/\mu}\right]^{3/5}
\left[\frac{d(t_n)}{M_p/\mu}\right]^{3/5}
\left[\frac{m}{10^{16} GeV}\right]^{-6/5},
\end{equation}
where we have assumed $\alpha \sim \gamma G\mu$ and followed the
argument in ref.\cite{branden}.
In this case, although the ``initial'' number density of monopoles that
could be contained in a loop increases with time, one may obtain $r\ll
1$ if annihilation is an efficient process.
To be precise, assuming efficient annihilation,
the actual number density of monopole per unit
length may become a constant(i.e. $r=const.$), which suggests that
 the ratio $r$ 
have an attractor point at $k=0$.\footnote{The
possibility that $r$ have an attractor is suggested in
\cite{vilenkin-necklace}.}

\section{PBHs from cycloops}
Here we follow the argument presented in ref.\cite{cycloops}.
According to ref.\cite{cycloops}, here we assume that there was a period
when strings can move freely in extra dimensions.
The number of black holes created  at time $t$ is given by
eq.(\ref{loop-production}), and the mass of black holes produced by a
chopped loop at $t$ is 
\begin{equation}
m(t)\sim \omega_l \mu \alpha t,
\end{equation}
where $\omega_l$ is a parameter that was used in ref.\cite{cycloops}.
For a loop of total length $l_{cy}$, its length in the compact dimensions
is given by $l_{cy}\omega_l$. 
It should be noted that before the ``lift'' of the potential one can use
the result that is obtained in ref.\cite{cycloops} for velocity
correlations. 
After the ``lift'', one can use the result obtained in our paper.
It is natural to think that the production of PBH starts much later than
the ``lift'' of the potential.
Then it is easy to obtain an upper bound\cite{branden};
\begin{equation}
\label{cycl}
G\mu < 10^{-15} \times
\left[\frac{p}{10^{-2}}\right]^{1/2}
\left[\frac{\gamma}{10^{2}}\right]^{-1/4}
\omega_l^{-3/4}
\end{equation}
where we have assumed $\alpha \sim \gamma G\mu$.

\section{Conclusions and Discussions}
The effective action of the daughter branes that are
produced by brane collision could have flat directions corresponding to
their free motion in the compactified space.
In this case, if the compactified manifold is not simply connected,
or valley of the potential for the string motion in extra dimensions has
a non-trivial circle,
stable winding states may appear.
Moreover, the effective potential that lifts the flat directions may
have degenerated vacua. 
Then, due to the random distribution of the ``vacua'', cosmic strings
will turn into necklaces and coils. 
In this paper, we have considered PBH formation from loops of cosmic
necklaces.
The existence of the winding states could be a serious problem.
Since the existence of the winding state in brane models is due to the
 existence of a 
 non-tivial circle in the compactified space, the PBH formation can be
used to probe the structure of the compactified space.
Black holes produced by this mechanism may have peculiar properties that
can be used to probe extra dimensions.

\section{Acknowledgment}
We wish to thank K.Shima for encouragement, and our colleagues in
Tokyo University for their kind hospitality.

\begin{appendix}
\section{From cycloops to necklaces-coils}
Here we make more comments about the scenario of
cycloops\cite{cycloops}, which should be helpful to 
understand why necklaces-coils are more important in our analysis.
In ref.\cite{cycloops}, cosmic loop production by long string
interactions in models with compact extra dimensions was studied,
where the important assumption is free motion in the extra dimensions.
In the case that the compact manifold is not simply connected,
there is a possibility that loops wrap around non-trivial circles.
The authors claimed that cycloops poses a potential monopole problem
because such loops behave like heavy matter in radiation era, and
calculated the energy density of cycloops.
Then they have shown that to avoid cycloop domination the strings must
satisfy the severe constraint $G\mu < 10^{-14}$, which may be consistent
with brane inflation.
In this scenario, however, the authors claimed that the mass of cycloops
increase with time as $m_{cycloops}\propto t^{1/2}$ if the strings obey
statistical model of random walk, and also that $m_{cycloops}\propto t$
if velocity correlations are considered. 
Therefore, one can easily understand that the PBH formation must be
crucial in the scenario of cycloops, because in this case
the production ratio of PBHs from loops cannot be suppressed, while the
mass increases with time as $m_{cycloops} \propto t$.
Here we should note that, even in the standard scenario of PBH formation
where the PBH production ratio is highly suppressed as $f < 10^{-20}$,
the upper bound 
obtained from PBH constraint is as low as $G\mu < 10^{-6}$.
Moreover, the present fraction of PBHs must satisfy
 (\ref{PBH_omega}), which is much stronger than $\Omega_{cycloops}<1$
 which has been considered in 
ref.\cite{cycloops}.
It is therefore important to understand PBH formation in the scenario of
cycloops.

\begin{figure}[ht]
 \begin{center}
 \begin{picture}(520,300)(0,0)
\resizebox{17cm}{!}{\includegraphics{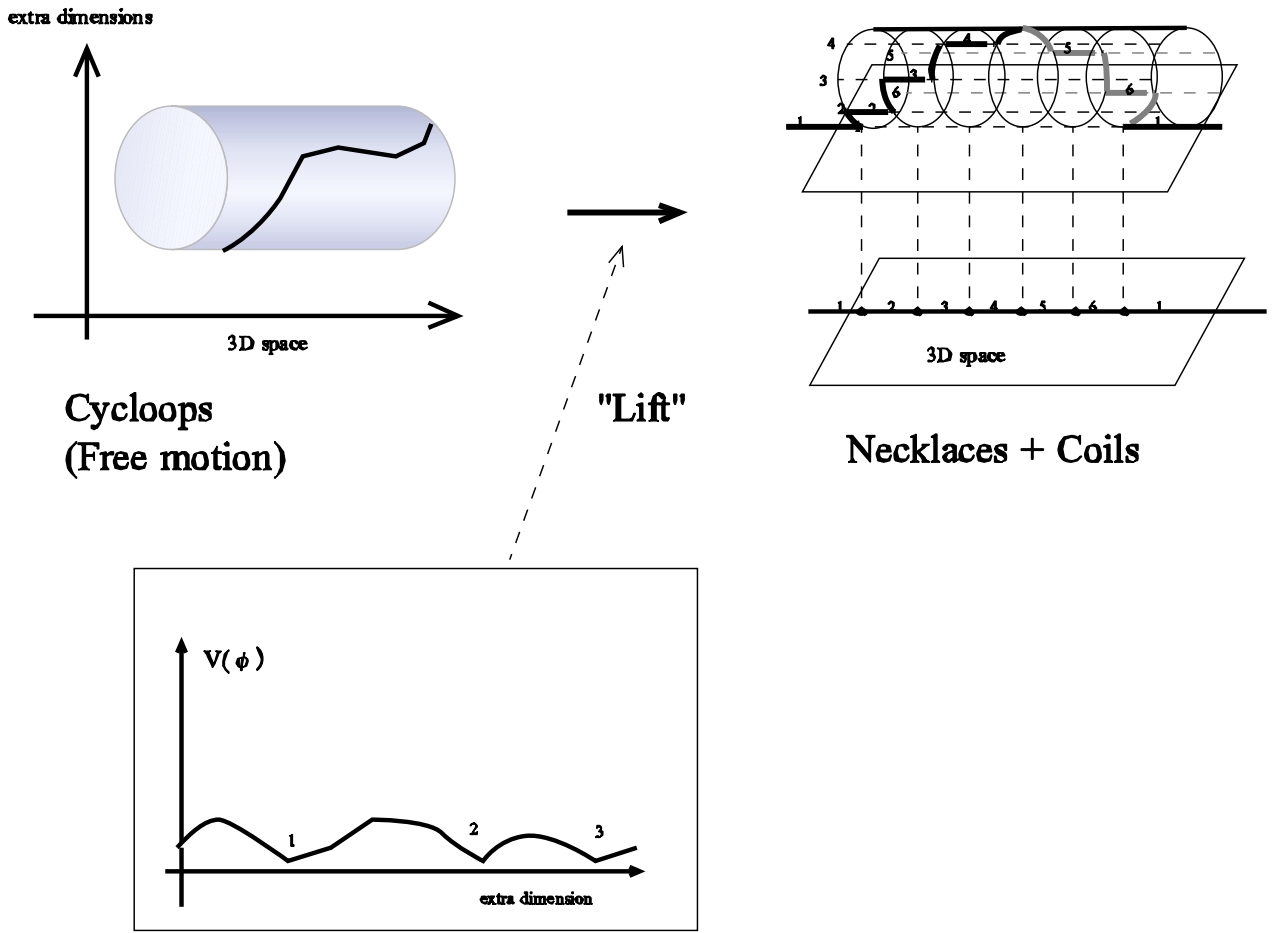}} 
 \end{picture}
\caption{This picture shows the relation between cycloops and
  necklaces-coils.}
\label{appendix}
 \end{center}
\end{figure}

However, since PBH production starts at late epoch, one
cannot assume free motion in the extra dimensions, but should consider
the ``lift'' from the potential.\footnote{The situation is shown in
fig.\ref{appendix}.}
This is the reason why one should consider necklaces/coils rather than
cycloops.\footnote{We will discuss dark matter production 
in our forthcoming paper\cite{matsuda_toap}.}
Here we should note that one cannot simply ignore the effective
potential that lifts the moduli, particularly the one lifts the flat
direction that corresponds to the string motion in the internal space.
If the potential is so high that the strings cannot move in the extra
dimensions, necklaces/coils are unlikely.
However, if strings can climb up the potential hill at least in the most
energetic epoch when strings are just produced, there could be random
distribution of ``vacua'' on the strings and kinks that
interpolate between them.

If the potential stabilizes the vacuum, one should consider cosmic
necklaces/coils instead of cycloops.
Therefore, here we consider the network of necklaces/coils
in scaling epoch, because the production of primordial black holes starts
late.  

\end{appendix}

\end{document}